**Title:** Interfacial ferroelectricity in rhombohedral-stacked bilayer transition metal dichalcogenides

**Authors:** Xirui Wang[1†], Kenji Yasuda[1†*], Yang Zhang[1], Song Liu[2], Kenji Watanabe[3], Takashi Taniguchi[3], James Hone[2], Liang Fu[1] and Pablo Jarillo-Herrero[1*]

**Affiliations:**

[1]Department of Physics, Massachusetts Institute of Technology, Cambridge, Massachusetts, USA.

[2]Department of Mechanical Engineering, Columbia University, New York, NY, USA.

[3]National Institute for Materials Science, Tsukuba, Japan.

[†]These authors contributed equally to this work

*Correspondence to: yasuda@mit.edu, pjarillo@mit.edu

**Introductory paragraph:**

Van der Waals (vdW) materials have greatly expanded our design space of heterostructures by allowing individual layers to be stacked at non-equilibrium configurations, for example via control of the twist angle. Such heterostructures not only combine characteristics of the individual building blocks, but can also exhibit emergent physical properties absent in the parent compounds through interlayer interactions[1]. Here we report on a new family of emergent, nanometer-thick, semiconductor 2D ferroelectrics, where the individual constituents are well-studied non-ferroelectric monolayer transition metal dichalcogenides (TMDs), namely $WSe_2$, $MoSe_2$, $WS_2$, and $MoS_2$. By stacking two identical monolayer TMDs in parallel, we obtain electrically switchable rhombohedral-stacking configurations, with out-of-plane polarization that is flipped by in-plane sliding motion. Fabricating nearly-parallel stacked bilayers enables the visualization of moiré ferroelectric domains as well as electric-field-induced domain wall motion with piezoelectric force microscopy (PFM). Furthermore, by using a nearby graphene electronic sensor in a ferroelectric field transistor geometry, we quantify the ferroelectric built-in interlayer potential, in good agreement with first-principles calculations. The novel semiconducting ferroelectric properties of these four new TMDs opens up the possibility of studying the interplay between ferroelectricity and their rich electric and optical properties[2–5].

**Main text:**

Symmetry plays a central role in the electronic band structure and the electronic properties of crystals. Recent advances in the fabrication of 2D layered materials and their heterostructures have enabled convenient control of their crystal symmetry and electronic structures[1]. One of the most prominent examples is the 2H-type transition metal dichalcogenides (TMDs), $MX_2$. Although the bulk crystal is inversion symmetric, spatial inversion symmetry is broken when it is exfoliated down to a monolayer, since the metal (M) and chalcogen (X) atoms occupy different crystallographic sites. The in-plane broken inversion symmetry, together with the spin-orbit coupling, endows monolayer TMDs with an interesting band structure characterized by spin-valley locking, where the spin direction is locked to the valley pseudospin direction[6]. Such spin-valley locking has been shown to induce various unique optical and electronic properties, *e.g.*, valley-polarized excitons[7], the valley Hall effect[8], and Ising superconductivity[9–11]. The symmetry and the electronic band structure of TMDs can be further modified by stacking the two monolayers with a precisely controlled angle. When two layers are stacked in an antiparallel manner, the hexagonal-stacked (H-stacked) bilayer TMD is realized, and global inversion symmetry is restored (Fig. 1a). In contrast, when two layers are stacked in parallel, the rhombohedral-stacked (R-stacked) structure is realized[12,13], which breaks the out-of-plane mirror symmetry, in contrast to monolayer, H-stacked bilayer, or bulk 2H-type crystals. The R-stacked structure takes either an MX- or XM-stacked form, where every M atom on the top layer lies over an X atom in MX stacking, while every X lies over an M in XM stacking (Fig. 1b,c)[14]. The broken mirror symmetry causes interlayer charge transfer through hybridization between the occupied states of one layer and the unoccupied states of the other layer, generating an out-of-plane electric dipole moment[2,3]. Since the two stacking orders are transformed into each other by a lateral shift of the two layers, the out-of-plane polarization could be switched by an in-plane interlayer shear motion, resulting in interfacial ferroelectricity[15,16].

The concept of interfacial ferroelectricity has been recently successfully demonstrated in parallel-stacked bilayer boron nitride (BN), where BN is turned into a ferroelectric when two sheets of BN are stacked in parallel[17–19]. In contrast to the conventional top-down approach to obtain 2D ferroelectrics by exfoliating layered polar materials[20–23], the bottom-up approach based on van der Waals assembly can engineer 2D ferroelectrics out of non-ferroelectric parent compounds[17–19,24]. As the latter approach is not limited by the thermodynamic stability of the bulk crystal, it can significantly expand the available 2D ferroelectric materials. In this paper, focusing on the similarity of the TMD and BN crystal structures, we generalize and identify four new interfacial ferroelectrics, namely R-stacked bilayer $WSe_2$, $MoSe_2$, $WS_2$, and $MoS_2$. In addition to the potential application for non-volatile memory devices[4,15,25], R-stacked bilayer TMDs can incorporate intrinsic ferroelectricity into semiconductors to realize ferroelectric semiconductor field-effect transistors[4,5]. Moreover, as R-stacked bilayer TMDs exhibit polarization-dependent excitonic responses, one can expect ferroelectric switching of the optical properties[2,3].

First, we performed PFM on a small-angle twisted bilayer $MoSe_2$ crystal at room temperature to characterize the physical responses on MX and XM domains. Since MX- and XM-stacking are the local energy minima in the parallel stacked form, small-angle twisted TMDs prepared by the tear-and-stack method[26,27] will form triangular domains composed of MX and XM stacking regions (after lattice relaxation[14]). In fact, triangular patterns of domains are observed in the vertical PFM amplitude image (Fig. 1e). The strong contrast at the domain wall between the adjacent domains (red dotted lines) is attributed to the flexoelectric effect originating from the strain gradient at the domain wall[28]. In addition, we find a relatively weak but finite contrast between the adjacent MX and XM domains (blue circles). As strain gradients are negligible inside the domain, the contrast purely originates from the piezoelectric effect, meaning that each domain has spontaneous out-of-plane polarization in the opposite direction. This observation is consistent with the previous reports in which broken mirror symmetry was detected as contrast in the tunneling current in conductive atomic force microscopy[14].

As MX and XM domains have opposite out-of-plane polarizations, they will react differently to the out-of-plane electric field. To study their responses, we performed lateral PFM on an R-stacked bilayer $MoSe_2$ under electric field, as shown in Fig. 1e. Since lateral PFM is sensitive to the flexoelectric effect at the domain wall[28], but not to the piezoelectric effect at the domain, we track the evolution of the domains through domain wall motion. At $V_B = -4$ V, up and down domains coexist in the field of view (Fig.1e, left schematic and adjacent image). As $V_B$ is changed from $-4$ V to 2 V, the domain walls gradually move, leading to the shrinkage of the down domains until they almost vanish (Fig.1e, right schematic and the adjacent image). The down domains expand when $V_B$ is swept back, confirming the ferroelectric nature of the R-stacked bilayer $MoSe_2$ (see Fig. S3 and Supplementary Video 1 for detailed images). We note that the polarization switching always occurs through domain wall motion, not through the nucleation of a domain within our electric field range. This indicates that ferroelectric switching may be harder to realize in single-domain devices, and points to the important role of pre-existing domain walls pinned by wrinkles, cracks, and bubbles for the switching.

The ferroelectricity of R-stacked bilayer TMDs is also detectable by using graphene as an electric sensor[18,23], which not only allows an accurate estimation of the polarization but also is significant as a proof-of-concept demonstration of a ferroelectric field-effect transistor. For these purposes, we fabricated dual-gated devices as shown in the inset of Fig. 2a, where the resistance of graphene reflects the ferroelectric polarization of the R-stacked bilayer TMDs. As a narrow resistance peak is ideal for probing the polarization, we sandwich graphene with top and bottom BN so as not to directly contact the TMDs, which prevents the broadening of the resistance peak due to spin-orbit coupling[29], disorder, and defects[30] from TMDs (see Fig. S4 and discussion in Supplementary Materials). Measuring the resistance as a function of the top gate does not lead to hysteresis (Fig. 2a), while prominent hysteresis is observed upon sweeping the bottom gate (Fig. 2b). The resistance peak at the charge neutrality point is shifted by $\Delta V_B/d_B = 0.018$ V nm$^{-1}$ between the forward and backward scan, where $V_B$ is the bottom gate voltage and $d_B$ is the total thickness of the bottom dielectrics. In addition, abrupt jump features of resistance are observed at $V_B/d_B = -0.35$ V nm$^{-1}$ and $V_B/d_B = +0.32$ V nm$^{-1}$, as shown in the inset of Fig. 2b. The appearance of hysteresis only in the bottom gate scan suggests that the hysteresis originates from the R-stacked bilayer $WSe_2$ between the bottom gate and graphene. Taken together with the PFM results, we interpret that the hysteresis and the jump features are due to ferroelectric switching through the change of the stacking configuration from MX to XM, or vice versa. The ferroelectric polarization induces additional charge carriers in the graphene, leading to the shift of the resistance peak as the electric field switches the polarization. To study the generality of the ferroelectricity in R-stacked bilayer TMDs, we applied the same procedure to $MoSe_2$, $WS_2$, and $MoS_2$ (Figs. 2c-e). The bottom gate scans exhibit two resistance peaks and the height of each peak changes depending on the sweeping direction. The appearance of the two resistance peaks can be explained by the coexistence of two domains, with each peak corresponding to each domain configuration. For example, the whole device is polarized upward in the backward scan of Fig. 2c, while MX and XM domains spatially coexist in the forward scan.

To further confirm the ferroelectric switching origin of the hysteresis, we measured the resistance of device WSe$_2$ d1 as a function of top and bottom gate voltages. We scanned the top gate repeatedly while changing the bottom gate slowly forward or backward, as indicated by the white arrows in Figs. 3a,b. We observe a diagonal feature corresponding to the graphene charge neutrality point resistance peak, as the bottom gate is scanned backward, with a sudden parallel shift of the diagonal line from the left to the right at $V_B/d_B = -0.28$ V nm$^{-1}$ (Fig. 3a) Similarly, a parallel shift from the right to the left is observed at $V_B/d_B = +0.37$ V nm$^{-1}$ for the forward scan (Fig. 3b). As in typical dual-gated devices, the diagonal feature tracks the charge neutrality point, where the net induced carrier density is approximately $n_0 = \varepsilon_{BN}(V_B/d_B+V_T/d_T)/e$, with $\varepsilon_{BN}$ being the dielectric constant of BN, $e$ being the elementary charge, and $d_T$ being the thickness of the top BN. The shift of the resistance peak is explained by the extra carrier density induced by the polarization of the R-stacked bilayer TMD: As the polarization switches from up to down, the total carrier density changes from $n_0+\Delta n_P$ to $n_0-\Delta n_P$. The coercive field in device WSe$_2$ d1 (~0.30 V nm$^{-1}$, Fig. 3c) is larger than those of parallel-stacked bilayer BN devices (~0.10 V nm$^{-1}$)[18], probably due the higher energy barrier between the oppositely polarized states and the smaller polarization in R-stacked bilayer TMDs (Fig. 4)[15,16].

Unlike the abrupt and complete switching in the entire device in WSe$_2$ d1, device WSe$_2$ d2 features step-like ferroelectric switching (Fig. 3d). We replot the resistance map in Fig. 3d as a function of $V_B/d_B+V_T/d_T$ and $V_B/d_B$ (Fig. 3e,f) to highlight the contribution of the extra charge carrier density induced by the polarization. Two vertical resistance peaks are observed in the skewed plots (Fig. 3e,f), which correspond to up and down domains, respectively. As the electric field $V_B/d_B$ is decreased (increased), the spectral weight shifts from the left (right) peak to the right (left) peak, meaning the expansion of down (up) domains. Hence, we can extract the average polarization $\langle P \rangle$ from the height of the two peaks by performing a two-peak fitting. We observe a multi-step change of polarization as a function of the electric field with a relatively small coercive field (Fig. 3c). Similar features are observed in all the devices except device WSe$_2$ d1 (Fig. S10-S13). We speculate that these behaviors originate from the pinning of multiple domain walls within a device. The ferroelectric switching due to the motion of pre-existing domain walls lead to a relatively small coercive field, and multi-step switching is induced by the independent motion of each domain wall.

Finally, we extract the built-in interlayer potential $\Delta V_P$ for each material and compare with theory. The blue line in the device schematic (Fig. 4a inset) represents the electrostatic potential profile across the graphite and graphene. When the polarization points up (down), an extra bottom gate voltage $V_B = -\Delta V_P$ ($V_B = +\Delta V_P$) is required to tune graphene to the charge neutrality point. Thus, we can extract $\Delta V_P$ from the relation $\Delta V_P = \Delta V_B/2$, where $\Delta V_B$ is the shift of the resistance peak or the separation of the two resistance peaks in the bottom gate scan (Fig. S16b). The experimentally obtained $\Delta V_P$ as a function of $d_B$ for each material is plotted in Fig. 4a. The variations of $\Delta V_P$ between each device of R-stacked bilayer $WSe_2$, $MoSe_2$, and parallel-stacked bilayer BN are all small, demonstrating the fidelity of our estimation. $\Delta V_P$ for the four R-stacked bilayer TMDs are all around 55 mV, which is about half of that in parallel-stacked bilayer BN[18]. The experimentally obtained built-in interlayer potentials are in reasonable agreement with our first-principles calculations, as shown in the Fig. 4b table. The consistency obtained for all the materials supports the reliability of both our experimental scheme and the theoretical calculations.

The ferroelectric hysteresis of R-stacked bilayer TMDs is robust up to room temperature (Figs. 3g and S14), and the switching operation at room temperature demonstrates the potential application for non-volatile memory devices (Fig. S15). The robust ferroelectricity in R-stacked bilayer TMDs not only vastly expands the families of ultrathin 2D ferroelectrics, but also opens up the possibility of studying the interplay between ferroelectricity and the abundant physical properties of TMDs as follows: First, as semiconductor TMDs are electrically doped by gating, we can obtain ferroelectric semiconductors where electronic conduction is switched on and off by the polarization direction[4,5]. Second, TMDs show extraordinary rich optical properties due to tightly bound exciton formation with valley degrees of freedom[2,3]. The ferroelectric polarization couples to these excitonic properties through band structure, enabling the non-volatile electrical control of the optical response. Third, beyond the parallel-stacked systems, the layer polarized nature of MX- and XM-stacked TMDs demonstrated in our study will serve as fundamental building blocks of twisted homo-bilayer TMDs, and will cause dramatic electric field response both in real and momentum spaces[14,31,32]. Finally, all of these unique properties will be easily incorporated by making heterostructures with different 2D materials to construct novel functional devices.

**Methods:**

Device fabrication:

$MoSe_2$ and $WSe_2$ crystals used for device $MoSe_2$ d2, $WSe_2$ d1-d3, and PFM measurement were grown by self-flux growth[33]. Crystals used for device $MoSe_2$ d1 and $WS_2$ were purchased from HQ Graphene. $MoS_2$ crystals were purchased from 2D Semiconductors. BN, graphite, and TMD crystals were exfoliated onto $SiO_2$ (285 nm-295 nm)/Si substrate. Graphene and monolayer TMDs were identified using optical contrast. The thickness of BN flakes was acquired by AFM. We exfoliated graphite for bottom gates on $SiO_2$ (285 nm)/Si substrates with pre-patterned markers, followed by heat cleaning in an atmosphere of Ar (40 sccm) and $H_2$ (20 sccm) gases at 350 °C for more than 12 hours to remove the tape residues. Each stack was made by the sequential pickup of top BN, graphene, bottom BN, and the bilayer TMD with a poly(bisphenol A carbonate) (PC)-film-covered Polydimethylsiloxane (PDMS) stamp on a glass slide. After picking up top BN, graphene, and bottom BN, we removed the trapped bubbles by moving the PC stamp and the stack up and down slowly a few times at 70-90 °C. R-stacked bilayer TMD is obtained by the sequential pickup of monolayer TMD flake at room temperature with the tear-and-stack method described in Refs 26 and 27. The whole stack was released to the bottom graphite at 170 °C. The stack was identified with an optical microscope and AFM for bubble-free regions. $MoS_2$ flakes were identified and transferred in a glovebox to avoid oxidation. The stack of $MoS_2$ was annealed in an Ar-only atmosphere at 70 sccm at 300 °C for 6 hours to reduce the bubbles. The stacks were etched into a Hall bar shape by reactive ion etching (RIE) for measurement, and all the contacts and top gates were deposited with Cr/Au with a thermal evaporator.

Piezoelectric force microscopy measurements:

The samples for vertical piezoelectric force microscopy (PFM) measurements (Figs. 1d and S2) were fabricated by picking up graphite and small-angle-twisted bilayer $MoSe_2$ with PC on PDMS in sequence. The PC films were peeled off from PDMS and placed onto SiO$_2$ (285 nm)/Si substrates, which were then heated up to 200 °C for better adhesion. For the sample for lateral PFM measurements (Fig. 1e), we first prepared BN (13.7 nm)/graphite stack, and heat cleaned in the atmosphere of Ar and H$_2$ gases at 300 °C for 14 hours. We then exfoliated MoSe$_2$ on PDMS and transferred small-angle-twisted bilayer MoSe$_2$. Finally, graphite exfoliated on PDMS was transferred to make electrical contact. The fabricated stack is annealed under Ar-only atmosphere at 170 °C for 10 mins. The PFM measurements were performed with Asylum Research Cypher S atomic force microscope at room temperature. We used Asyelec-01-R2 with a force constant of around 2.8 N m$^{-1}$ and a contact resonance frequency of around 280 kHz with the applied AC bias voltage of 1 V for the vertical PFM measurements. We used AC240TM-R3 with a force constant of around 1.5 N m$^{-1}$ and a contact resonance frequency of around 600 kHz with the applied AC bias voltage of 2 V for the lateral PFM measurements. The gate voltage was applied by a source meter (Keithley: Model 2400). The contact strength was set to be lower than 30 nN to avoid unintentional damage to the flake and twist angle relaxation.

Transport measurements:

The devices were bonded using aluminum wire. The four-probe measurements were done with lock-in amplifiers (SRS: SR830 and SR860), a current preamplifier (DL: Model 1211) and a voltage preamplifier (SRS: SR560) at a frequency of 35.5 Hz. The lock-in time constant was 100ms. The measuring current was 100 nA for device WSe$_2$ d1, WSe$_2$ d2, MoSe$_2$ d2, MoS$_2$, and 1 µA for device WSe$_2$ d3 and WS$_2$. The measurement of device MoSe$_2$ d1 was performed at 100 nA in the main text and 1 µA in the Supplementary Materials. The hysteresis measurements were performed by waiting for 1.5 s at each point to avoid measurement lag. The gate voltages were applied by source meters (Keithley: Model 2400). Devices WSe$_2$ d2 and WSe$_2$ d3 were measured in a He-3 cryostat (Janis research). Devices MoSe$_2$ d1, MoSe$_2$ d2, WS$_2$, and MoS$_2$ were measured in a homemade insert. Device WSe$_2$ d1 was measured in both systems. The data of device WSe$_2$ d1 in Fig. 3a,b were taken during the warm-up, when the temperature increased by 30 K due to the lack of helium. The temperature noted in Fig. 3a,b is the average temperature during this measurement.

DFT calculations:

Density functional calculations are performed using Perdew-Burke-Ernzerhof generalized gradient approximation[34] with the vdW correction incorporated by DFT-D3 method with Becke-Jonson damping[35], as implemented in the Vienna Ab-initio Simulation Package[36]. All the calculations are done with projector augmented wave (PAW) pseudopotentials with spin-orbit interaction included. The interlayer distance of R-stacked bilayer $MoSe_2$, $WS_2$, and $MoS_2$ is taken from the bulk lattice constant of 3R-TMDs, and the interlayer distance of R-stacked bilayer $WSe_2$ is taken from the lattice constant of bulk $2H\text{-}WSe_2$ as an approximation[12,37,38]. The vacuum size was chosen to be 20 Å to simulate isolated bilayers. To ensure the periodic boundary condition in the out-of-plane direction, we used the inverted double-bilayer structures and compensated the built-in electric field. The plane-wave cutoff energy is set to be 1200 eV to ensure the convergence of electrostatic potential inside the vacuum, and the Brillouin zone sampling mesh is 12×12×1. We calculated the sum of ionic potential and Hartree potential from charge density to evaluate the built-in interlayer potential. The calculation shows good consistency with an independent work on R-stacked bilayer TMDs[39].

**Acknowledgements:**

We thank Sergio de la Barrera for fruitful discussions. This research was primarily supported by the U.S. Department of Energy, Office of Science, Basic Energy Sciences, under Award Number DE-SC0020149 (measurement, data analysis, and DFT calculation), by the Center for the Advancement of Topological Semimetals, an Energy Frontier Research Center funded by the U.S. Department of Energy Office of Science, through the Ames Laboratory under contract DE-AC02-07CH11358 (device concept and design), by the Army Research Office (nanofabrication development) through grant #W911NF1810316, and the Gordon and Betty Moore Foundations EPiQS Initiative through grant GBMF9643 to P.J-H. This work made use of the Materials Research Science and Engineering Center Shared Experimental Facilities supported by the National Science Foundation (NSF) (Grant No. DMR-0819762). This work was performed in part at the Harvard University Center for Nanoscale Systems (CNS), a member of the National Nanotechnology Coordinated Infrastructure Network (NNCI), which is supported by the National Science Foundation under NSF ECCS award no. 1541959. K.W. and T.T. acknowledge support from the Elemental Strategy Initiative conducted by the MEXT, Japan, Grant Number JPMXP0112101001, JSPS KAKENHI Grant Numbers JP20H00354 and the CREST(JPMJCR15F3). K.

Y. acknowledges partial support by JSPS Overseas Research Fellowships. Synthesis of WSe$_2$ and MoSe$_2$ was supported by the NSF MRSEC program through Columbia in the Center for Precision-Assembled Quantum Materials (DMR-2011738).

**Author contributions:**

K.Y. designed and conceived the project. X.W. and K.Y. fabricated the devices and performed transport measurements. K.Y. performed PFM measurements. Y.Z. and L.F. performed the theoretical calculation. S.L. and J.H. grew the WSe$_2$ and MoSe$_2$ (used in device MoSe$_2$ d1) crystals. K.W. and T.T. grew the BN crystal. X.W., K.Y, Y.Z., and P.J-H. analyzed the data and wrote the manuscript with the input from all the other authors.

**Competing interests:**

The authors declare no competing interests.

Figures:

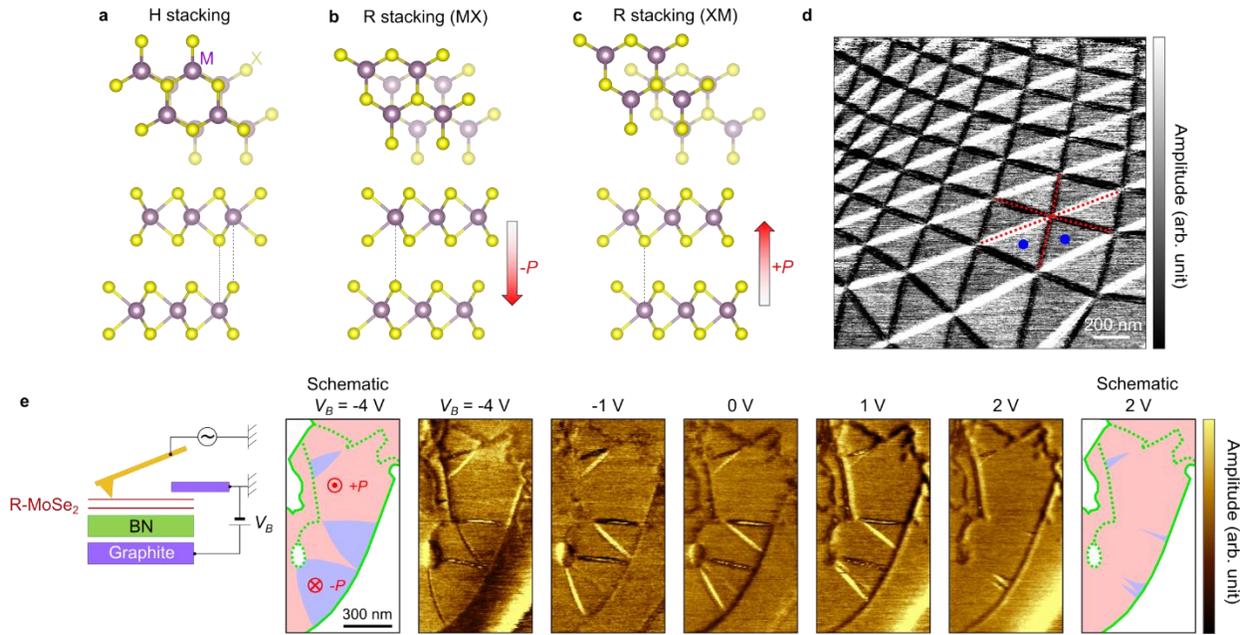

Figure 1: Crystal structures and piezoelectric force microscopy of bilayer TMDs. **a**, H-stacked bilayer TMD. The two layers are stacked in antiparallel configuration, and the inversion symmetry is restored as a whole. M: metal atom (W or Mo). X: chalcogen atom (S or Se). **b, c,** MX (**b**) and XM (**c**) stacking forms of R-stacked bilayer TMD. The two layers are stacked in parallel, and an out-of-plane polarization exists due to the vertical alignment of different atoms. **d**, Amplitude image of vertical PFM on small-angle-twisted bilayer $MoSe_2$. The adjacent triangular domains (blue dots) exhibit finite PFM contrast. **e**, Schematic illustration of lateral PFM measurement on a gated small-angle-twisted bilayer $MoSe_2$ device (left). The thickness of BN is 13.7 nm. Amplitude image of lateral PFM under different bottom gate voltages $V_B$. Schematic of the domain configuration is illustrated for $V_B = -4$ V (left) and $V_B = 2$ V (right), as a guide to the eyes. The area surrounded by the solid green curves is the bilayer $MoSe_2$ region. The wrinkles, cracks, and bubbles are shown in dotted green curves. The red and blue regions correspond to up and down domains, respectively.

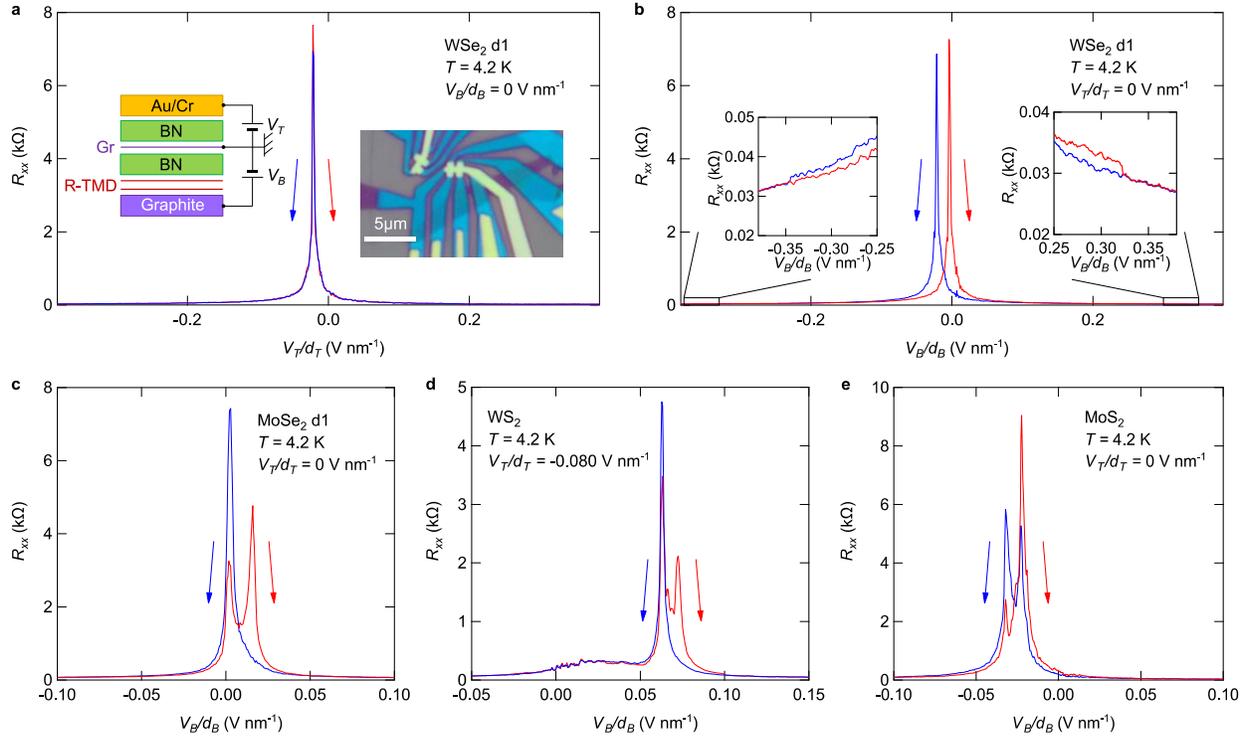

Figure 2: Hysteresis in R-stacked bilayer TMD devices. **a**, The resistance of graphene in device WSe$_2$ d1 as a function of top gate in the forward (red) and backward (blue) scan directions. Scan range: −0.43 V nm$^{-1}$ to +0.51 V nm$^{-1}$. Inset: The device structure and the optical image of device WSe$_2$ d1. The data was taken from the device on the right. **b-e**, The resistance of graphene in R-stacked bilayer TMD devices as a function of bottom gate in the forward and backward scan directions. The scan ranges in **c-e** are enlarged for a better visualization of the resistance peaks. **b**, WSe$_2$ d1. Scan range: −0.38 V nm$^{-1}$ to +0.38V nm$^{-1}$. Inset: enlarged data. **c**, MoSe$_2$ d1. Scan range: −0.22 V nm$^{-1}$ to +0.28 V nm$^{-1}$. **d**, WS$_2$. Scan range: −0.12 V nm$^{-1}$ to +0.15 V nm$^{-1}$ with −0.080V nm$^{-1}$ applied to the top gate. **e**, MoS$_2$. Scan range: −0.17 V nm$^{-1}$ to +0.19 V nm$^{-1}$.

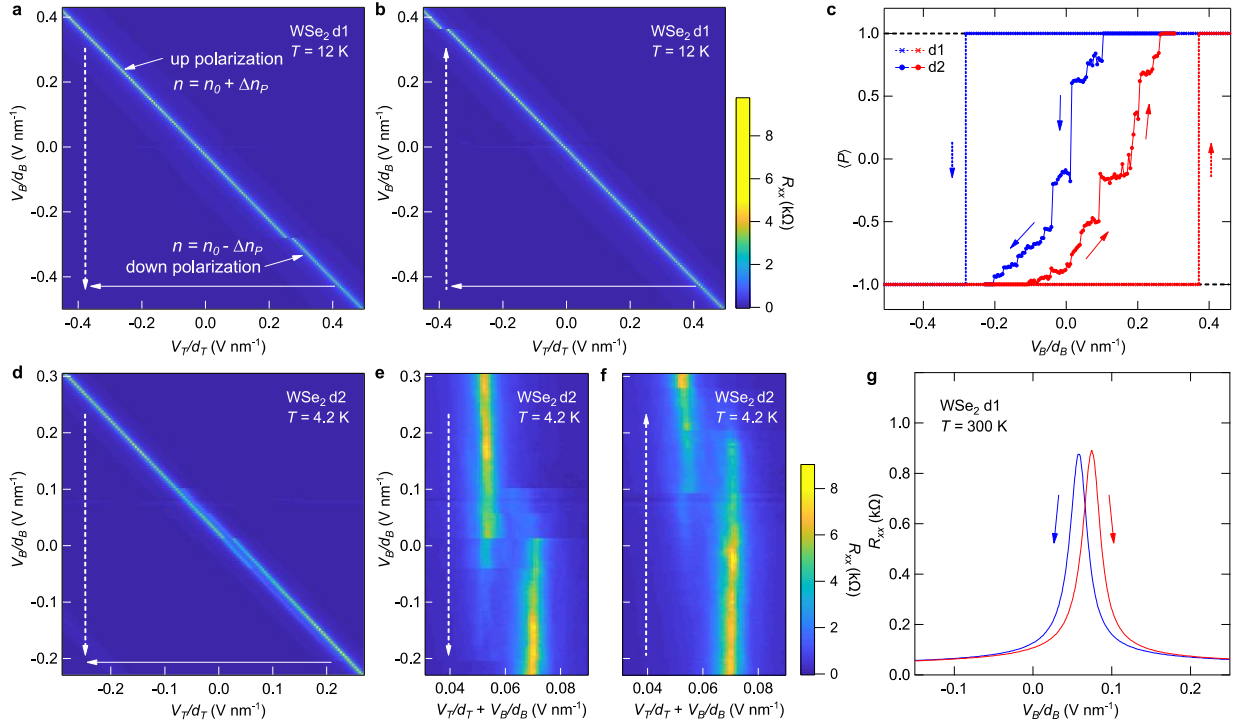

Figure 3: Electric field dependence of the polarization. **a**, Bottom gate backward scan of device WSe$_2$ d1. The fast-scan axis, $V_T/d_T$, was scanned from $+0.62$ V nm$^{-1}$ to $-0.47$ V nm$^{-1}$. The slow-scan axis, $V_B/d_B$, was scanned from $+0.46$ V nm$^{-1}$ to $-0.51$ V nm$^{-1}$. The two parallel-shifted resistance ridges correspond to up and down domains, with carrier densities equal to $n_0+\Delta n_P$, and $n_0-\Delta n_P$, respectively ($n_0 = \varepsilon_{BN}(V_B/d_B+V_T/d_T)/e$). The ranges in the $x$ and $y$ axes are slightly enlarged for a better visualization of the resistance peak, same for **b** and **d**. **b**, Bottom gate forward scan of device WSe$_2$ d1. **c**, Average polarization $\langle P \rangle$ in device WSe$_2$ d1 and device WSe$_2$ d2 as a function of $V_B/d_B$ in the forward (red) and backward (blue) scan directions. $\langle P \rangle$ of device WSe$_2$ d2 was acquired by two-peak Lorentzian fitting of the linecuts of the dual-gate scan. **d**, Bottom gate backward scan of device WSe$_2$ d2. The fast-scan axis, $V_T/d_T$, was scanned from $+0.32$ V nm$^{-1}$ to $-0.29$ V nm$^{-1}$. The slow-scan axis, $V_B/d_B$, was scanned from $+0.30$ V nm$^{-1}$ to $-0.23$ V nm$^{-1}$. **e**, Skewed plot of **d** with $V_B/d_B+V_T/d_T$ as the $x$ axis. **f**, Skewed plot of the bottom gate forward scan with the same scan range as **d**. **g**, The resistance of graphene in device WSe$_2$ d1 as a function of bottom gate in the forward (red) and backward (blue) scan directions at $T = 300$ K. Scan range: $-0.38$ V nm$^{-1}$ to $+0.46$ V nm$^{-1}$.

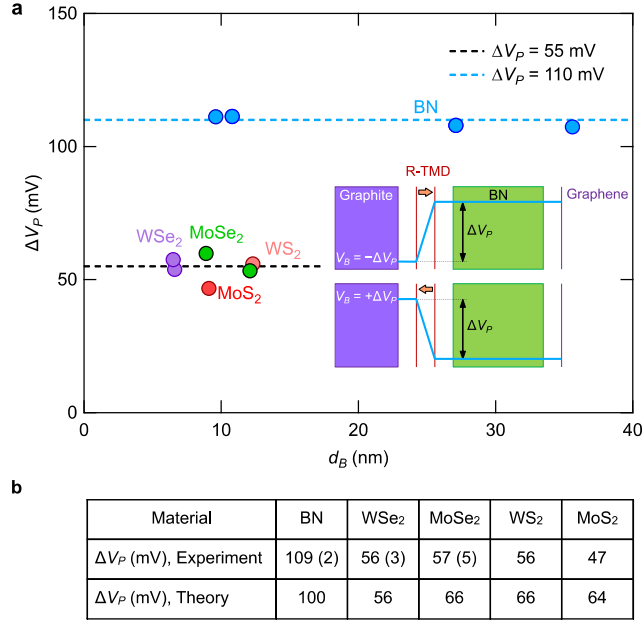

Figure 4: Estimation of built-in interlayer potential in R-stacked bilayer TMDs and comparison with theory. **a,** Built-in interlayer potential, $\Delta V_P$, plotted as a function of the total thickness of bottom dielectrics, $d_B$. Parallel-stacked bilayer BN (extracted from Ref. 18) (blue), WSe$_2$ (purple), MoSe$_2$ (green), MoS$_2$ (red), and WS$_2$ (pink). The experimental value of BN was extracted from the original data in Ref. 18. Dashed lines represent $\Delta V_P = 55$ mV (black) and $\Delta V_P = 110$ mV (blue) as a guide to the eyes. Inset: Schematic of electrostatic potential profile between graphite and graphene. Blue line: electrostatic potential. The upper and lower figures show the opposite polarized states. **b,** Experimental and theoretical estimations of $\Delta V_P$ in parallel-stacked bilayer BN and R-stacked bilayer TMDs. The numbers in the parentheses indicate the standard deviation between the devices.